   \documentclass[prl,aps,twocolumn,showpacs]{revtex4}
\usepackage[dvips]{graphicx}
\usepackage{dcolumn}
\usepackage{amsmath}
\usepackage{amsfonts}
\usepackage{amssymb}
\providecommand{\U}[1]{\protect\rule{.1in}{.1in}}
\newcommand{\be}{\begin{equation}}
\newcommand{\ee}{\end{equation}}
\newcommand{\bea}{\begin{eqnarray}}
\newcommand{\eea}{\end{eqnarray}}
\newcommand{\nn}{ \nonumber}
\newcommand{\ds}{\displaystyle}
\begin{document}

\title{The electron transport through a quantum dot in the Coulomb blockade regime: 
Non-equilibrium Green's functions based model}

\author{ Natalya A. Zimbovskaya}

\affiliation{Department of Physics and Electronics, University of Puerto 
Rico-Humacao, CUH Station, Humacao, PR 00791,\\  Institute for Functional 
Nanomaterials, University of Puerto Rico, San Juan, PR 00931;
 \\   Institute for Nanoelectronics and Computing, School of 
Electrical and Computer Engineering, \\ Purdue University,
465 NW Ave., West Lafayette, IN 47907}

\begin{abstract} 
 We explore electron transport through a quantum dot coupled to the source and 
drain charge reservoirs. We 
trace the transition from the Coulomb blockade regime to Kondo regime in the 
electron transport through the dot occuring when we gradually strengthen the 
coupling of the dot to the charge reservoirs.  
The current-voltage 
$(I-V)$ characteristics are calculated using the equations of motion approach 
within the nonequilibrium Green's functions formalism (NEGF) beyond the
Hartree-Fock approximation. 
We show that within the Coulomb blockade regime the 
$I-V$ characteristics for the quantum dot containing a single spin-degenerated 
level with the energy $E_0 $ include two steps whose locations are determined 
by the values of $ E_0 $ and the energy of Coulomb interaction of electrons in 
the dot $U.$ The heights of the steps are related as $2:1$ which is consistent 
with the results obtained by means of the transition rate equations. 
\end{abstract}

\maketitle
\date{\today}

\section{i. Introduction: }

As known, quantum dots host such interesting quantum transport phenomena as 
Coulomb blockade \cite{1} and Kondo effect \cite{2,3}. Fast development of 
experimental technique in recent years provides broad opportunities for 
fabrication of devices including novel active elements such as single molecules, 
nanowires and carbon nanotubes, placed in between metallic leads. Such junctions 
are commonly treated as quantum dots coupled to the macroscopic charge reservoirs 
which play the parts of  source and drain in the electron transport. Current 
achievements in nanoelectronics make it possible to observe electron transport 
through realistic junctions  varying external parameters (source-drain and gate 
voltage and temperature) and internal characteristics of the junction, e.g. the 
coupling strength of the quantum dot to the source and drain reservoirs.  As a 
result, both supression of  the electron transport through the quantum dot at 
small values of the source-drain voltage $(V)$ arising due to the high charging 
energy inside the dot (Coulomb blockade), and the increase in the electrical 
conduction of the junction at $ V \approx 0 $ due to the Kondo effect were 
repeatedly observed in experiments on molecular  and carbon nanotubes junctions 
\cite{4,5,6,7,8,9,10,11,12,13,14}.

The Coulomb blockade occurs when  the quantum dot representing a molecule or a 
carbon nanotube is weakly coupled to the source and drain which means that the 
charging energies $U $ are much greater than the contact and temperature induced 
broadening of the electron levels in the dot.
Theoretical studies of the electron transport through a quantum dot within the 
Coulomb blockade regime mostly employ ``master" equations for the transition 
rates between the states of the dot differing by a single electron 
\cite{15,16,17,18}. It was shown that the current-voltage curves for a dot 
including a single spin-degenerated level exhibite two steps. The steps 
correspond to the gradual adding/removing two electrons with the different 
spin orientations to/from the dot. Assuming the symmetric contacts of the dot 
with both charge reservoirs, the first step in the current-voltage curve must 
be two times higher than the second one (see Refs. \cite{18,19}). This result 
has a clear physical sense. Obviously, one may put the first electron of an 
arbitrary spin orientation to the empty level in the dot. However, the spin of 
the second electron added to the dot is already determined by the spin of the 
first one. So, there are two ways to add/remove the first electron to/from the 
empty/filled level in the dot but only one way to add/remove the second one. 

While the approach based on the transition rates ``master" equations brings 
sound results within the Coloumb blockade regime, its generalization to the 
case of stronger coupled junctions is not justified. More sophisticated methods 
employing various modifications of the nonequilibrium Green's functions  
formalism (NEGF) are developed to analyze electron transport  through  
 quantum dots coupled to the leads  \cite{20}. 
 These theoretical techniques were repeatedly employed to calculate the 
retarded/advanced and lesser Green's functions for the dot. The Green's functions 
are needed to compute the electron density of states (DOS) on the dot where 
the Kondo peak could be manifested, as well as the electron current through the junction. 
A commonly used approach within the NEGF is the equations of motion (EOM) method 
which was first proposed to describe quantum dot systems by Meir, Wingreen and 
Lee \cite{21}. This technique enables to express the relevant Green's functions 
in terms of higher-order Green's functions. Writing out EOM for the latter, one 
arrives at the infinite sequence of the equations successively involving Green's 
functions of higher orders.

To get expressions for the 
necessary Green's functions this system of EOM must be truncated. Also, 
higher-order Green's functions still included in the remaining EOM must be 
approximated  to express them by means of the lower-order Green's functions. 
In outline, this procedure is well-known and commonly used. However, in numerous 
existing papers important details of the above procedure vary which brings 
different approximations  for the relevant Green's functions. Accordingly, 
the results of theoretical studies   of the quantum dot response vary.

The NEGF formalism was successfully applied to describe the quantum dot response 
within the Coulomb blockade regime (weak coupling of the dot to the electron 
reservoirs), using the expressions for both retarded and lesser Green's function 
within the Hartree-Fock approximation \cite{22,23}. The results for the electron 
transport through the quantum dot obtained in these works give the correct ratio 
of the subsequent current steps, namely: $ 2:1, $ assuming the dot to be 
symmetrically coupled to the leads.
It is a common knowledge that within the 
Hartree-Fock approximation one cannot catch the Kondo peak. To 
quantitatively describe this effect one may need to include  hybridization 
up to very high orders in calculation of the lead-dot coupling parameters, and 
the studies aiming at better theoretical analysis of the Kondo effect started 
by Meir et al \cite{24} are still in progress (see. e.g. Refs. \cite{20,25,26}). 

Obviously, the Green's functions suitable to reveal the Kondo peak in the 
electron DOS on the dot must provide the proper description of the 
transport  through the dot within the Coulomb blockade regime when leads-dot 
couplings are weak.
 However, usually  NEGF based results  beyond the Hartree-Fock approximation
successfully employed to describe the Kondo effect 
and related phenomena, were not applied to the limiting case of the Coulomb 
blockade. In those works were such application was carried out, the correct ratio 
of heights of the subsequent current steps, namely $2: 1, $ was not obtained. 
 For instance, Muralidharan, Ghosh and Datta \cite{18} 
reported the results on NEGF calculations which give equal heights of the two 
steps on the current-voltage curve within the Coulomb blockade regime, and 
Galperin, Nitzan and Ratner studies \cite{20} result in the heights ratio 
about $ 1.6 : 1. $
So, up to present there exists a discrepancy between NEGF and master equations 
based results. 
This discrepancy may not be easily 
disregarded for its existence makes questionable general and useful results 
based on the advanced techniques within the NEGF formalism. Recently, the above 
disagreement was discussed in the Ref. \cite{18}.

The purpose of the present work is to clarify the issue, and to erase the above 
disconsistency. We apply equation of motion based approach to NEGF method to 
get a description of both Coulomb blockade and Kondo regimes in the electron 
transport through a quantum dot 
 within the same computational procedure for the relevant Green's functions.
We show that the assumed approximations give 
correct ratio of heights of the steps in the current-voltage curves within the 
Coulomb blockade limit and allow to get the Kondo peak, as well. 
For simplicity, we omit from consideration dissipative 
effects, and we concentrate on a non-magnetic quantum dot 
coupled to non-ferromagnetic leads. Obtained results 
 are valid for an arbitrary value of the Coulomb interaction on the dot, 
assuming that the latter is stronger than dot-leads coupling. Also, 
the analysis may be generalized to magnetic systems and to
 dots containing many levels provided that the charging energy 
is small compared to the adjacent levels separations. 
 We remark that the present analysis is confined with using 
simplest possible approximations for the necessary Green's functions. Therefore 
we do not take into account corrections proposed in Ref. \cite{26} to better 
approximate the Kondo peak.

\section{ii. Main equations} 

We write the Hamiltonian of the junction including the dot and source and drain 
reservoirs (leads) as:  
  \be 
H = H_L + H_R + H_D + H_T.
  \label{1} \ee
  Here, the terms $H_\beta\ (\beta =L,R) $ corrrespond to the left and right 
leads and  describe them within the noninteracting particles approximation
  \be
  H_\beta = \sum_{r\sigma} \epsilon_{r\beta\sigma} 
c_{r\beta\sigma}^{+} c_{r\beta\sigma} 
  \label{2} \ee
  where $\epsilon_{r\beta\sigma}$ are the single-electron energies in the 
electrode $ \beta $ for the electron states $r,\sigma $ (the index 
$ \sigma $ labeling spin up and 
spin down electrons), and $c_{r\beta\sigma}^{+}, c_{r\beta\sigma} $ denote 
the creation and annihilation operators for the electrodes. 
 We assume that energy levels $ \epsilon_{r\beta\sigma} $ in the electron 
reservoirs are uniformly spaced over the ranges corresponding to the electron 
conduction bands for both spin orientations. The spacing between two adjacent 
levels $ \delta \epsilon_{\beta\sigma} $ is supposed to be much smaller than 
the bandwidth $ \Delta_{\beta\sigma}\ (\delta\epsilon_{\beta\sigma}/
\Delta_{\beta\sigma} \sim 10^{-4}).$

The term $H_D $ describes the dot itself. Taking into account the 
electron-electron interaction, this term has the form:
  \be
 H_D = \sum_\sigma E_\sigma c^+_\sigma c_\sigma + U c_\uparrow^+ 
c_\uparrow c_\downarrow^+ c_\downarrow
   \label{3} 
  \ee 
  where $c_\sigma^+,c_\sigma $ are the creation and annihilation operators 
for the electrons in the dot;  $E_\sigma = E_0 $ is the energy of the 
single dot level, the energy $U $ characterizes the Coulomb interaction 
of electrons in the dot. We carry out calculations within the low temperature 
regime, so $U$ is supposed to significantly exceed the thermal energy $kT.$ 
The last term in the Hamiltonian Eq. \ref{1} 
describes the coupling of the reservoirs to the dot:
  \be 
  H_T = \sum_{r\beta\sigma} \tau_{r\beta\sigma}^* c_{r\beta\sigma}^+ 
c_\sigma +H.c.
  \label{4} \ee
  where $\tau_{r\beta\sigma} $  are the coupling parameters
describing the coupling of the $ r,\sigma $ electron states belonging to the
electrode $ \beta $ to the quantum dot.

Now, we employ the equations of motion method 
 to compute the retarded Green's function for the dot.
 Since no spin-flip processes on the dot are taken into account, and we assume 
the electron transport to be spin-conserving, we may separately introduce the 
Green's functions for each spin channel.
  The computational procedure is 
described in the Appendix. We obtain:   
  \be 
G_\sigma^{rr} (E) = \frac{E - E_0 -\Sigma_{02}^\sigma - U(1 - 
\big<n_{-\sigma}\big>)}{(E - E_0 -\Sigma_{0\sigma})(E - E_0 - U -
\Sigma_{02}^\sigma) + U \Sigma_{1\sigma}},
  \label{5} \ee
   Here, $\big<n_{-\sigma}\big> = \big<c_{-\sigma}^+ c_{-\sigma}\big>, $ 
 and self energy corrections are given by:  
   \begin{align} 
\Sigma_{0\sigma} =& \sum_{r\beta} \frac{|\tau_{r\beta\sigma}|^2}{E - 
\epsilon_{r\beta\sigma} + i\eta}, \label{6}
   \\
 \Sigma_{1\sigma} = & \sum_{r\beta}|
\tau_{r\beta,-\sigma}|^2 f_{r,-\sigma}^\beta
 \left \{\frac{1}{E -2E_0 - U + 
\epsilon_{r\beta,-\sigma} + i\eta}\right.  
  \nn\\ & +\left.
  \frac{1}{E - \epsilon_{r\beta,-\sigma} + i\eta} \right\},\label{7}
  \\
\Sigma_{2\sigma} = & \sum_{r\beta} |\tau_{r\beta,-\sigma}|^2 
 \left \{\frac{1}{E -2E_0 - U + \epsilon_{r\beta,-\sigma} + i\eta}\right.
  \nn\\ & +\left.
  \frac{1}{E - \epsilon_{r\beta,-\sigma} + i\eta} \right\},\label{8}
  \\ 
\Sigma_{02}^\sigma = & \Sigma_{0\sigma} + \Sigma_{2\sigma}. \label{9}
 \end{align}
   Here, $ f_{r\sigma}^\beta $ is the Fermi distribution  function for the 
energy $ \epsilon_{r\beta\sigma} $ and the chemical potential $ \mu^\beta $ 
and $ \eta $ is a positive infinitesimal parameter. The expression \ref{5} for 
the Green's function $ G_\sigma^{rr} (E)$ was first derived by Meir and 
Wingreen \cite{21}. Later, the same expression was obtained in Ref. \cite{25} 
and (assuming that the electron energy levels on the leads are spin degenerate) 
in the book \cite{19}. It was repeatedly used to qualitatively describe the Kondo 
peak in the electron DOS (see e.g. Refs. \cite{24,25}). As shown 
in these works, the Kondo peak  originates from the last term in the denominator 
of the Eq. \ref{5}. At low temperatures  the real part of the self-energy term 
$\Sigma_{1\sigma}$ diverges giving rise  to the peak at $ \mu^L = \mu^R = E. $ 
Better approximations for the Green's function $G_\sigma^{rr} (E)$ such as that 
derived in the Ref. \cite{26} may give better results for the shape and height 
of the Kondo peak. However, the very emergence of the latter is accounted for in
the Eq. \ref{5}.

The occupation numbers $ \big<n_\sigma\big> = \big<c_\sigma^+ c_\sigma\big> $ and 
$ \big<n_{-\sigma}\big> = \big<c_{-\sigma}^+ c_{-\sigma}\big> $ may be found 
solving the integral equations of the form:
   \be
\big<n_\sigma\big> = \frac{1}{2\pi} \int dE \mbox{Im} (G_{\sigma}^<)
  \label{10}  \ee
   where the lesser Green's fucntion $G_{\sigma}^<$ is introduced. 
 The function $G_{\sigma}^< $ obeys the Keldysh 
equation:  
\be 
  G_\sigma^< (E) = G_\sigma^{rr} (E) \Sigma_\sigma^< G_\sigma^{aa} (E)\label{11}
  \ee
 where $ G_\sigma^{aa} (E) $ is the Fourier component of the advanced Green's 
function.
The Fourier transform of the  lesser Green's function $G_\sigma^< $ is 
related to the Fourier transforms of the retarded and advanced Green's 
functions of the form \ref{5}, so the quantities $\big< n_\sigma\big> $ and 
$\big< n_{-\sigma}\big> $ appear in the integrands in the right hand sides of 
the equations \ref{10} for both spin directions. 
In further calculations we use the following expression for 
$ \Sigma_\sigma^<:$
  \be 
  \Sigma_\sigma^< = i \sum_\beta f_\sigma^\beta (E) \Gamma_\sigma^\beta.\label{12}
   \ee
 Consequently, we have
   \be
  G_{\sigma}^< = iG_\sigma^{rr} (E) G_\sigma^{aa} (E) 
\left[\Gamma_\sigma^L f^L_\sigma(E) + \Gamma_\sigma^R f^R_\sigma (E)\right]
  \label{13}  \ee
  where $f^{L,R}_{\sigma} (E)$ are the Fermi distribution functions for the 
left/right reservoirs, and
    \be
 \Gamma_{\sigma}^\beta = - 2 \mbox{Im} \left(\Sigma_{0\sigma}^\beta\right)
  \label{14}  \ee
 where 
  \be \Sigma_{0\sigma}^\beta = \sum\limits_{r} 
\ds\frac{|\tau_{r\beta\sigma}|^2}{E -\epsilon_{r\beta\sigma} +i\eta}. 
  \label{15} \ee

 Employing the expression \ref{12} for $\Sigma_\sigma^< $ we assume that the 
Coulomb interaction of the electrons on the dot does not affect the coupling 
of the latter to the leads, as well as it occurs within the Hartree-Fock 
approximation \cite{22}. Such assumption agrees with the way of estimating 
averages in the process of decoupling of high-order Green's functions which 
was used to derive Eq. \ref{5}. It seems reasonable while the dot coupling to the 
leads is weaker than the characteristic energy of the Coulomb interactions on 
the dot. The proper choice of approximation for the lesser Green's function is 
very important for it directly affects the values of occupation numbers 
$\big<n_{\pm\sigma}\big> $ and, consequently, the heights of the peaks in the 
electron DOS in the Coulomb blockade regime. Also, the relative 
heights of the steps in the current-voltage characteristics depend on the 
occupation numbers. The employed form for $ G_\sigma^< (E) $ differs form those 
used in some previous works (see e.g. Refs. \cite{20,25}) where  extra terms 
arising due to the Coulomb interaction on the dot are inserted in the expression 
for $ \Sigma_\sigma^<.$ Certainly, one may expect such terms to appear as 
corrections to the main approximation \ref{12}. However, these terms must become 
insignificant and negligible  in the Coulomb blockade regime when the dot is 
weakly coupled to the leads. The expression for $ \Sigma_\sigma^< $ reported 
in Ref. \cite{25} does not satisfy this requirement.

Substituting Eq. \ref{5} into 
Eq. \ref{13} and inserting the result into Eq. \ref{10}
 we obtain the system of equations to find the mean occupation numbers of 
electrons on the dot:
  \be
 \big< n_{\pm\sigma}\big>  = P_{\pm\sigma} + U \big<n_{\mp\sigma}\big> 
Q_{\pm\sigma} + U^2 \big< n_{\mp\sigma}\big>^2 R_{\pm\sigma} .  
   \label{16} \ee
  Here,
 \begin{align}
 & P_{\pm\sigma} =\frac{1}{2\pi} \sum_\beta \int \Gamma_{\pm\sigma}^\beta 
f_\sigma^\beta(E) A_{\pm\sigma}  A_{\pm\sigma} ^+ dE, \label{17}
\\ 
 & Q_{\pm\sigma} =  \frac{1}{2\pi} \sum_\beta \int \Gamma_{\pm\sigma}^\beta 
f_\sigma^\beta(E) \left(A_{\pm\sigma}  B_{\pm\sigma} ^+ + A_{\pm\sigma}^+  
B_{\pm\sigma}\right) dE,\label{18}
 \\ 
 & R_{\pm\sigma} = \frac{1}{2\pi} \sum_\beta \int \Gamma_{\pm\sigma}^\beta 
f_\sigma^\beta(E) B_{\pm\sigma}  B_{\pm\sigma} ^+ dE,  \label{19}
  \end{align}
 and the expressions for $A_\sigma $ and $B_\sigma $ are obtained using 
the Green's function \ref{5}, namely: 
   \begin{align} 
 A_\sigma = \frac{E - E_0 - U -\Sigma_{02}^\sigma}{(E - E_0 -\Sigma_{0\sigma})
(E - E_0 - U -\Sigma_{02}^\sigma)+ U\Sigma_{1\sigma}},
  \nn\\ \nn\\
 B_\sigma = \frac{1}{(E - E_0 -\Sigma_{0\sigma})(E - E_0 - U -\Sigma_{02}^\sigma)
+ U\Sigma_{1\sigma}}. \label{20}
  \end{align}

  When considering a nonmagnetic system, the energy levels 
in the source and drain (and in the quantum dot, as well) are spin degenerated, 
 so $ A_\sigma = A_{-\sigma},\ 
B_{\sigma}=B_{-\sigma}. $Correspondingly the occupation numbers for both 
spin orientations coincide with each other. In this case $ \big<n_\sigma\big> = 
\big<n_{-\sigma}\big> $ are described with the equation
   \be
 \big<n_\sigma\big> = \frac{1 - UQ  - \sqrt{(1- UQ)^2 - 4U^2RP}}{2U^2 R}.
  \label{21} \ee
  When the electron correlations on the quantum dot are weak $(U\to 0),$ 
the result \ref{21} is reduced to $ \big<n_\sigma\big> = P_\sigma. $
 
The Eq. \ref{21} is an important result for it gives an explicit analytical 
expression for the level populations, so we do not need a time-consuming 
self-consistent iterative procedure to find the occupation numbers \cite{27}.  
Now, we write the expression for the electric current $I$ flowing through the 
junction when we apply the voltage $ V $ across the latter. This expression was 
derived by Jauho, Wingreen and Meir (see refs. \cite{28,29}), and it has the 
form:
   \bea
I &=& \frac{ie}{2\hbar} \sum_\sigma \int \frac{dE}{2\pi} 
\left[\left(\Gamma_\sigma^L f_\sigma^L - \Gamma_\sigma^R f_\sigma^R\right) 
\left(G_\sigma^{rr} - G_\sigma^{aa} \right) \right.
   \nn\\ && \left.
+ \left(\Gamma_\sigma^L - \Gamma_\sigma^R \right) G_\sigma^< \right]
 . \label{22}  \eea  
   Also, we could employ the equivalent expression:
   \bea
I &=& \frac{ie}{2\hbar} \sum_\sigma \int \frac{dE}{2\pi} 
\left[\left(\Gamma_\sigma^L f_\sigma^L - \Gamma_\sigma^R f_\sigma^R\right) 
\left(G_\sigma^{rr} - G_\sigma^{aa} \right) \right.
   \nn\\ && \left.
- i\left(\Gamma_\sigma^Lf_\sigma^L + \Gamma_\sigma^R f_\sigma^R\right) 
G_\sigma^{rr} G_\sigma^{aa} 
\left(\Gamma_\sigma^L - \Gamma_\sigma^R \right) \right].
   \label{23} \eea
   Further we assume the symmetrical voltage division: $ \mu_{L,R} = E_F \pm 
\frac{1}{2} eV, $ and we put $ E_F = 0. $ Also, in further calculations we 
introduce a gate potential $ V_g, $ so that 
  \be 
 E_0 (V_g) = E_0 (V_g =0) + eV_g.
  \label{24} \ee
  The application of the gate voltage shifts the energy level in the dot, so 
the energies $E_0 $ and $E_0 + U $ become asymmetrically arranged with respect 
to $ E = E_F. $ This asymmetry is necessary to obtain two steps in the $I-V $ 
curve assuming the symmetric voltage division. Otherwise, the contributions to 
the current from the vicinities of $ E = E_0 $ and $E = E_0 + U $  become 
superposed, and only one step emerges. 

\section{iii. Discussion}

To simplify further calculations we assume 
all coupling strengths to take on the same value, namely: 
$ \tau_{r\beta\sigma} = \tau.$
Now, we calculate the current through the junction in the limit of the weak 
coupling of the dot to the reservoirs $(\tau \ll U).$ To carry out the 
calculations we need to know the nonequilibrium occupation numbers and we 
compute them using Eq. \ref{21}. The occupation numbers are sensitive to the
value of the applied voltage at small voltages as shown in the Fig. 1. 
At low values of the bias voltage $(V < 0.4V)$ the assumed dot energy level 
$(E_0 = - 0.2eV)$ is situated below the chemical potentials for both leads. 
Therefore the dot is able to receive an electron but unable to transfer it to 
another lead. Accordingly, the average occupation number is close to unity and 
the electric current through the junction takes on values close to zero (see 
Fig. 2). At $ V = 0.4 V $ the dot energy $ E_0 $ crosses the chemical potential 
$ \mu^R,$ and the dot becomes active in electronic transport. Now, the electron 
which arrive at the dot from one reservoir may leave it for another reservoir. 
This results in a pronounced decrease in the average occupation on the dot 
accompanied by an increase in the current. One more change in both average 
electron occupation on the dot and the current through the junction occurs at 
$ V = 0.6 V$ when the energy $ E_0 + U\ (U = 0.5 eV) $ crosses $ \mu^R. $ 
At higher voltage 
all curves presented in the figures 1,2 level off. The current reachs its maximum 
value and the average number of electrons in the dot reachs its minimum. The 
minimum occupation number is noticeably less than one but its value is nonzero 
for electrons unceasingly travel through the junction.

\begin{figure}[t]   
\begin{center}
\includegraphics[width=7.2cm,height=9.2cm,angle=-90]{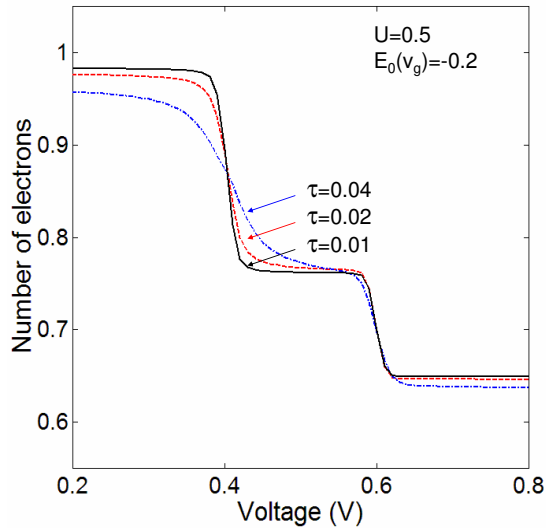}
\end{center}
\caption{
 Average occupation numbers in the quantum dot as function of 
the voltage applied across the junction.
 The curves are plotted assuming $ U = 0.5 eV,\ 
kT=0.00026 eV,\ E_0(V_g) = -0.2 eV,\  \tau = 0.01 eV $ (solid line), $ 0.02 eV $ 
(dash-dotted line) and $0.04 eV$ (dashed line). }
\label{rateI}
\end{figure}

\begin{figure}[t]
\begin{center}
\includegraphics[width=7.2cm,height=9.5cm,angle=-90]{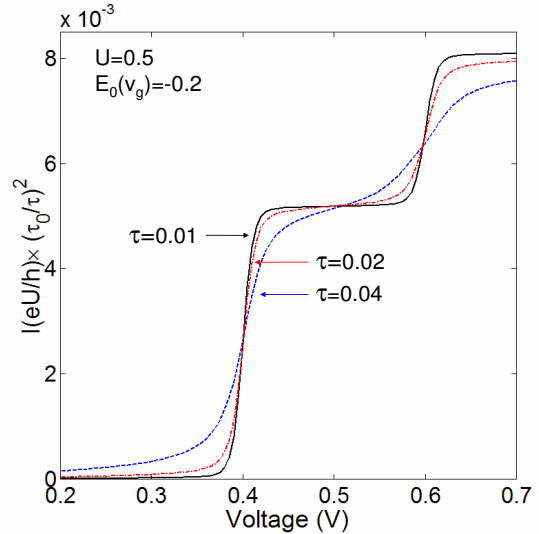}  
\end{center}
\caption{
Current through a junction within the Coulomb blockade 
regime. The curves are plotted assuming $  U = 0.5 eV,\ kT=0.00026 
eV,\ E_0(V_g) = -0.2eV,\
\ \tau_0 = 0.04 eV,\  \tau = 0.01 eV $ (solid line), $ 0.02 eV $ (dash-dotted 
line) and $0.04 eV$ (dashed line). The factor $ \tau_0^2 /\tau^2 $ is introduced 
to bring all $ I-V $ curves to the same scale. The ratio of heights of the two 
steps revealed in the curves equals $2:1. $}
\label{rateI}
\end{figure}

The current-voltage curves in the Fig. 2
 show typical Coulomb blockade features, namely, two steps whose 
heights are related as $2:1. $ So, we see that the results concerning the 
electron transport through a quantum dot obtained employing the transition 
rates equations may be quantitatively reproduced within the NEGF formalism 
when the retarded and lesser Green's functions are approximated by Eqs. \ref{5}, 
\ref{13}, and the explicit expression \ref{21} for the occupation 
numbers is used. Therefore, the disagreement discussed in the begining of the 
present work may be successfully erased, and the consistency between the NEGF 
formalism and the transitions rates equations in the description of the Coulomb 
blockade regime could be restored  beyond the Hartree-Fock approximation.

It is worthwhile to remark that particular values of the average occupation
numbers on the dot are very responsive to the gate voltage value $ V_g$   
(the latter determines how the dot energy level is situated with  respect to 
the Fermi levels of the leads in the absence of the voltage applied across them)
and to the Coulomb interaction energy $U.$ Therefore, different values chosen for 
$V_g$  and $U $ lead to different  average occupation numbers.
However, under various assumptions for the $V_g$ and $ U $ values, one quantity
does not vary provided the symmetric coupling of the dot to the leads. 
This quantity is the relative height of the subsequent steps in 
the average occupation numbers of the electrons in the dot which are revealed 
as the voltage across the leads increases. As shown in the Fig. 1 this ratio 
is $2:1, $ exactly the same as for the subsequent current steps in the $I-V $ 
curves in the Fig. 2. And it is this ratio which ensures the correct shape of
$I-V $ curves.  For 
instance, comparing figure 1 with the corresponding result reported on the
work \cite{22}, one may see that the values of the occupation numbers given
in Ref. \cite{22} considerably differ from those obtained in the present work.
Nevertheless, the ratio of the subsequent steps heights in the occupation 
numbers versus voltage curves is $ 2:1 $, and this provides for the same ratio of
heights of the subsequent current steps.

\begin{figure}[t] 
\begin{center}
\includegraphics[width=7.5cm,height=9.5cm,angle=-90]{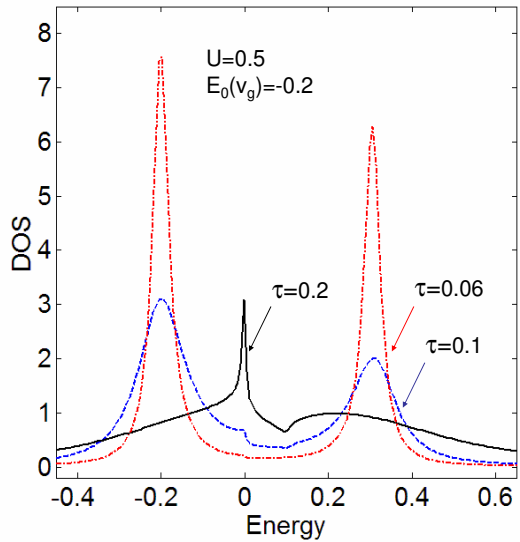} 
\end{center}
\caption{
The equilibrium  $(V=0)$ electron density of states in the quantum dot. The 
curves are plotted for $\tau = 0.2 eV $ (solid line) $ 0.1 eV $ (dashed line) 
and $ 0.06 eV $ (dash-dotted line). The remaining parameters are the same as 
those in the figure 1.}
\label{rateI}
\end{figure}
 
\begin{figure}[t] 
\begin{center}
\includegraphics[width=7.5cm,height=9.5cm,angle=-90]{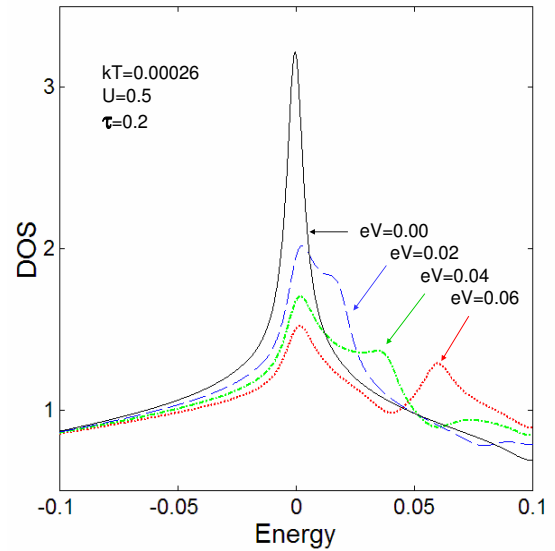} 
\end{center}
\caption{
Splitting of the Kondo peak in the electron density of states with increasing 
bias. The curves are plotted for $ eV = 0 $ (solid line); $ 0.02 eV$ (dashed 
line), $0.04 eV $ (dash-dotted line) and $ 0.06 eV $ (dotted line). The coupling 
strength $ \tau = 0.2 eV. $ The remaining parameters are taken as in the figures 
1,2. }
\label{rateI}
\end{figure}

Also, the Eqs. \ref{5}, \ref{13}, \ref{19} could be applied to analyze 
the Kondo effect which is manifested at stronger coupling strengths of the dot 
to the charge reservoirs. The electron density of states (DOS) in the 
dot is given as 
  \be
 D(E) = - 2 \sum_\sigma \mbox{Im} \left[G_\sigma^{rr} (E)\right].
  \label{25} \ee
 Using $G_\sigma^{rr} (E)$ in the form \ref{5} we may compute the electron 
DOS. The results are presented in the Fig. 3 where the equilibrium 
DOS is shown for three values of the coupling strength $ \tau. $ For a 
sufficiently strong coupling of the dot to the source and  drain 
reservoirs $(\tau = 0.2 eV) $ the sharp Kondo peak appears at $ E = 0, $ 
and the peaks at $ E = E_0 $ and $ E = E_0 + U $ are damped. At weaker 
coupling strength $(\tau = 0.1 eV,\ 0.07eV) $ the Kondo peak is reduced 
to a tiny feature but the maxima at $ E = E_0 $ and $ E = E_0 + U $ which 
determine the conductance within the Coulomb blockade regime
emerge. The weaker is the coupling the higher become these peaks.

The heights of the peaks differ. Technically, this distinction originates 
from the fact that the value of $\big< n_{-\sigma} \big> $ in the expression 
for the retarded Green's function (5) differs from $0.5. $ As shown in the book 
\cite{19}, assuming $ \big<n_{-\sigma} \big> = \big<n_\sigma\big> = 0.5 $ 
one would get the Coulomb blockade 
peaks of equal height but such assumption leads to the wrong result for the 
current. Namely, one would obtain two steps of equal height on the current voltage 
curves. The physical reason of this difference in the peak heights is the same 
as for the difference in heights of the steps on the current-voltage 
characteristics within the Coulomb blockade regime. The latter was discussed 
in the Introduction. 
When the voltage is applied across the junction the Kondo peak splits in two 
maxima whose heights are significantly smaller than the height of the original 
equilibrium Kondo peak and the greater is the voltage  the lower these maxima 
become. This is shown in the Fig. 4.   
So, the present formalism sufficiently reproduces the results of earlier works 
concerning the Kondo effect (see e.g. Refs. \cite{24,25}).

Finally, in the present work we theoretically analyzed the electron transport 
through a single quantum dot coupled to the source and drain charge reservoirs. 
The analysis was based on the NEGF formalism. 
 The expression (5) for the retarded Green's function was obtained 
using the equations of motion method, and agrees with the results of the previous 
works \cite{21,24,25}. The lesser Green's function was found from the Keldysh 
equation using the factor $ \Sigma_\sigma^< $ in the form (12). This means that 
$ \Sigma_\sigma^< $ is supposed to be unchanged due to the Coulomb interactions 
on the quantum dot, as it is proved to be within the Hartree-Fock approximation. 
We believe the proposed approximation to be appropriate at moderate and/or weak 
coupling of the dot to the leads when the coupling strengths are smaller than 
the characteristic Coulomb energy on the dot. 
We derived an 
explicit expression for the occupation numbers of electrons in the dot (Eq. 
\ref{21}) which enables us to compute them avoiding the long iterative 
procedure. The employed formalism gives correct results in the Coulomb blockade 
regime corroborating the results following from the transition rates (master) 
equations. Our approach is able to quantitatively reproduce the relative heights 
$(2:1)$ of the steps in the current voltage curves and to qualitatively 
describe the Kondo peak using the same expressions for the relevant Green's 
functions.
To the best of our knowledge 
this was never achieved  so far. 

Also, we remark that the present work addresses a general 
problem arising within the NEGF based computations. To quantitatively describe 
fine features of Kondo effect and related phenomena one needs to include 
hybridization up to very high orders in calculations of the relevant Green's 
functions. Carrying out these cumbersome calculations where numerous 
approximations are inserted, it is easy to make a mistake and difficult to 
discover one, especially in the absence of obvious small parameters. Probably, 
such  a mistake would not affect the very existence of the Kondo peak but it 
could distort its fine features. In the present work we propose a method which 
could help to verify the resulting expressions for the Green's functions. We 
start from the clear point, namely, that the Green's functions suitable to 
describe the Kondo effect are suitable to describe the Coulomb blockade transport,
 as well. Therefore, one may verify obtained results by going to the Coulomb 
blockade limit. If it occurs that particular Green's functions fail to provide 
the correct form of the I-V curves within this limit then one must conclude that 
there is some error in the adopted approximations for the Green's functions.

 It is shown that calculational scheme employed in the 
present work which uses a very simple approximation for $ G_\sigma^< $ brings 
quantitatively  correct results for the electron transport through a quantum dot 
within the Coulomb blockade regime whereas advanced self-consistent calculations 
 carried out in the recent paper \cite{20} do 
not. This gives grounds to conjecture that it is not necessary (and, perhaps, it 
is not always correct) to use the same number of iterations in seeking
approximations for retarded/advanced and the lesser Green's functions within EOM
method. Also, we remark that validity of the Green's functions used in studies
of the Kondo effect may be verified by applying them to calculate electron
transport within the Coulomb blockade regime.
The results could be 
generalized to include more realistic case of a dot including many electron 
levels, assuming that level separations are much greater than the Coulomb
energy $U,$ so one may neglect Coulomb interactions of electrons belonging
to different levels.

\section{ Acknowledgments:}
Author thanks S. Datta for helpful discussions, and G. M. Zimbovsky for help 
with the manuscript. This work was supported by DoD grant W911NF-06-1-0519 and 
NSF-DMR-PREM 0353730.

\section{appendix}

Here, we explain how the expression (5) for the Green's function for the dot is 
derived. We introduce the following notation
for a general retarded Green's function Fourier component:
  \be 
 \big<\big< A,B\big>\big> \equiv - i \int_0^\infty 
\big<\{A(t); B\} \big > e^{i(E + i\eta)t} dt
   \ee
  where $A,B$ are operators, 
curly brackets denote the anticommutator and $\big <\cdots\big>  $ stands for the 
average. For the Hamiltonian (1) the equation of motion for the retarded Green's 
function for the dot $G_\sigma^{rr} (E) \equiv \big<\big< c_\sigma; c_\sigma^+ 
\big>\big> $ reads:
   \begin{align} 
 (E -E_0 + i\eta) \big<\big< c_\sigma; 
c_\sigma^+ \big>\big> =
1 & + \sum_{r,\beta} \tau_{r\beta,\sigma}
\big<\big< 
c_{r\beta\sigma}; c_\sigma^+ \big>\big>
 \nn\\ & +
 U \big<\big< n_{-\sigma} 
c_\sigma; c_\sigma^+ \big>\big>   .\label{27}
  \end{align}
  Here, $n_{-\sigma} \equiv c_{-\sigma}^+ c_{-\sigma}, $ and the Green's function 
$ \big<\big< c_{r\beta\sigma} ; c_\sigma^+ \big>\big> $
obeys the equation:
  \be 
 (E - \epsilon_{r\beta\sigma} + i\eta) 
\big<\big< c_{r\beta\sigma} ; 
c_\sigma^+ \big>\big> = \tau_{r\beta\sigma}^*
\big<\big< c_{\sigma} ; 
c_\sigma^+ \big>\big> . \label{28}
   \ee
  Substituting the expression for $ \big<\big< 
c_{r\beta\sigma} ; c_\sigma^+ \big>\big> $ determined by \ref{28} into 
\ref{27} we get: 
   \be 
  (E - E_0 - \Sigma_{0\sigma}) 
\big<\big< c_{\sigma} ; 
c_\sigma^+ \big>\big> = 1 +
U \big<\big< n_{-\sigma} c_{\sigma} ; c_\sigma^+ 
\big>\big> \label{29}
  \ee
  where the self-energy part $ \Sigma_{0\sigma} $ is given by 
Eq. \ref{6}. The equation of motion for the four-operator function $ \big<\big< 
n_{-\sigma} c_{\sigma} ; c_\sigma^+ \big>\big> $
includes higher order Green's 
functions. Omitting them we write out:
 \begin{widetext} \begin{align}  
 & (E - E_0 - U + i\eta) 
\big<\big< n_{-\sigma} c_{\sigma} ; c_\sigma^+ \big>\big> 
  \nn\\ 
= & \big<n_{-\sigma} \big>  +  \sum_{r,\beta} \Big( \tau_{r\beta\sigma}
\big<\big<  c_{r\beta\sigma} n_{-\sigma}; c_\sigma^+ \big>\big> + 
\tau_{r\beta;-\sigma}
\big<\big< c_{-\sigma}^+ c_{r\beta;-\sigma} c_\sigma; c_\sigma^+ \big>\big>
-\tau_{r\beta;-\sigma}^* \big<\big<  c_{r\beta;-\sigma} c_{-\sigma} c_{\sigma}; 
c_\sigma^+ \big>\big> \Big). \label{30}
   \end{align}
  To proceed we must write equations for the Green's functions inserted in the 
right side of the Eq. \ref{30}. We get:
  \begin{align}  
 & (E - \epsilon_{r\beta\sigma} + i\eta) 
\big<\big<  c_{r\beta\sigma} n_{-\sigma}; c_\sigma^+ \big>\big> 
   \nn\\
= & \tau_{r\beta\sigma}^*
\big< \big< n_{-\sigma}  c_{\sigma} ; c_\sigma^+ \big>\big> +
 \sum_{r'\beta} \big( \tau_{r'\beta;-\sigma}
\big<\big< c_{r\beta\sigma} c_{-\sigma}^+ c_{r'\beta;-\sigma}; 
c_\sigma^+ \big>\big> 
- \tau_{r'\beta;-\sigma}^*
\big<\big< c_{r'\beta-\sigma} c_{-\sigma} c_{r\beta;\sigma}; 
c_\sigma^+ \big>\big> \big),\label{31}
 \\ 
 & (E - \epsilon_{r\beta;-\sigma} + i\eta )
\big<\big< c_{-\sigma}^+ c_{r\beta;-\sigma} c_\sigma ;c_\sigma^+  \big>\big>
 \nn\\ 
=  &  \tau_{r\beta;-\sigma}^* 
\big<\big< n_{-\sigma} c_\sigma; c_\sigma^+ \big>\big>  -
\sum_{r'\beta} \big(\tau_{r'\beta-\sigma}
\big<\big< c_{r'\beta;-\sigma}^+ c_{r\beta;-\sigma} c_\sigma; c_\sigma^+\big>\big>
  +\tau_{r'\beta\sigma}
\big<\big< c_{-\sigma}^+ c_{r\beta;-\sigma} c_{r'\beta\sigma}; 
c_\sigma^+ \big>\big> \big), \label{32}
 \\  
 & (E - 2E_0 - U + \epsilon_{r\beta;-\sigma} + i\eta) 
\big<\big<c_{r\beta;-\sigma}^+ c_{-\sigma} c_\sigma; c_\sigma^+ \big>\big>
  \nn\\ 
= & -\tau_{r\beta;-\sigma} 
\big<\big< n_{-\sigma} c_\sigma; c_\sigma^+ \big>\big>  
+ \sum_{r'\beta} \big (\tau_{r'\beta\sigma}
\big<\big<c_{r'\beta;-\sigma}^+ 
c_{-\sigma} c_{r\beta\sigma}; c_\sigma^+ \big>\big> 
- \tau_{r'\beta;-\sigma}
\big<\big< c_{r\beta;-\sigma}^+ c_\sigma c_{r'\beta;-\sigma}; 
c_\sigma^+ \big>\big> \big). \label{33}
   \end{align}   \end{widetext}
 Writing out these equations \ref{31}-\ref{33} we neglected averages like 
$\big<c_\sigma^+ c_{r\beta\sigma} \big>$ including a creation/annihilation 
operator for the dot combined with the annihilation/creation operator for the 
source/drain reservoir. Such averages are omitted in further calculations, 
as well. Now, we decouple four-operator Green's functions included in the sums 
over $``r'"$ in the equations following the scheme \cite{26}:   
 \be 
 \big<\big< A^+BC; D^+\big>\big> \approx 
\big< A^+ B\big> \big<\big< C; D^+ \big>\big> -
\big< A^+ C\big> \big<\big< B; D^+ \big>\big>. \label{34}
  \ee
  Also, we use the approximation:
  \be 
\big< c_{r\beta\sigma}^+ c_{r'\beta\sigma}\big> =
\big< c_{r'\beta\sigma} c_{r\beta\sigma}^+\big> =
\delta_{rr'} f_{r\sigma}^\beta  \label{35}
  \ee 
 where $ f_{r\sigma}^\beta $ is the Fermi distribution function corresponding 
to the energy $ \epsilon_{r\beta\sigma}.$ As a result, the Green's functions 
included in the right-hand side of the Eq. \ref{30} get expressed in terms of the 
Green's functions $\big<\big< n_{-\sigma} c_\sigma; c_\sigma^+ \big>\big>$ and 
$ \big<\big< c_\sigma; c_\sigma^+\big>\big>.$ Substituting these expressions 
into Eq. \ref{30} we obtain: 
  \be 
 \big<\big< n_{-\sigma} c_\sigma; c_\sigma^+ \big>\big> =
\frac{\big<n_{-\sigma}\big> + \Sigma_{1\sigma} 
\big<\big<c_\sigma;c_\sigma^+  \big>\big>}{E - E_0 - U - \Sigma_{02}^\sigma}. 
\label{36}
  \ee
   Here, self-energy parts 
$ \Sigma_{1\sigma}, \Sigma_{2\sigma}, \Sigma_{02}^{\sigma}$ are given by:
    \begin{align} 
 \Sigma_{1\sigma} = & \sum_{r\beta}|
\tau_{r\beta,-\sigma}|^2 f_{r,-\sigma}^\beta
 \left \{\frac{1}{E -2E_0 - U + 
\epsilon_{r\beta,-\sigma} + i\eta}\right.  
  \nn\\ & +\left.
  \frac{1}{E - \epsilon_{r\beta,-\sigma} + i\eta} \right\},\label{37}
  \\
\Sigma_{2\sigma} = & \sum_{r\beta} |\tau_{r\beta,-\sigma}|^2 
 \left \{\frac{1}{E -2E_0 - U + \epsilon_{r\beta,-\sigma} + i\eta}\right.
  \nn\\ & +\left.
  \frac{1}{E - \epsilon_{r\beta,-\sigma} + i\eta} \right\},\label{38}
  \end{align} \be  
\Sigma_{02}^\sigma =  \Sigma_{0\sigma} + \Sigma_{2\sigma}. \label{39}
   \ee 
  To arrive at the resulting expression for the dot Green's function 
$ G_\sigma^{rr} \equiv \big<\big< c_\sigma; c_\sigma^+ \big>\big> $ we 
substitute Eq. \ref{36} into Eq. \ref{29}. We have:
 \be 
G_\sigma^{rr} (E) = \frac{E - E_0 -\Sigma_{02}^\sigma - U(1 - 
\big<n_{-\sigma}\big>)}{(E - E_0 -\Sigma_{0\sigma})(E - E_0 - U -
\Sigma_{02}^\sigma) + U \Sigma_{1\sigma}}.  \label{40} 
    \ee

\end{document}